\documentclass[linenumbers]{aastex631}

\usepackage{amsmath,amsfonts,amssymb}

\newcommand{\blue}{\textcolor{black}}

\nolinenumbers

\shorttitle{Cross-scale energy transfer}
\shortauthors{Shi et al.}
\graphicspath{{./}{figures/}}

\begin{document}

\title{Energy transfer from MHD-scale slow-mode waves to kinetic-scale ion acoustic waves}

\correspondingauthor{Xiaofei Shi}
\email{sxf1698@g.ucla.edu}

\author[0000-0003-3367-5074]{Xiaofei Shi}
\affiliation{Department of Earth, Planetary, and Space Sciences, University of California, Los Angeles, CA, 90095, USA}

\author[0000-0003-2507-8632]{Xin An}
\affiliation{Department of Earth, Planetary, and Space Sciences, University of California, Los Angeles, CA, 90095, USA}



\author[0000-0001-7024-1561]{Vassilis Angelopoulos}
\affiliation{Department of Earth, Planetary, and Space Sciences, University of California, Los Angeles, CA, 90095, USA}




\begin{abstract}

\nolinenumbers

Large-amplitude slow-mode waves are commonly observed near Earth's magnetopause. Recent observations show that these waves can occur simultaneously with kinetic-scale ion acoustic waves. The amplitude of the ion acoustic waves is enhanced near the magnetic field peaks of the slow-mode wave, suggesting that the slow-mode waves may drive the generation of ion acoustic waves. To test this hypothesis, we conduct a hybrid simulation using observation-based parameters. The simulation results demonstrate that large-amplitude slow-mode waves generate counter-streaming ion beams, which in turn excite ion acoustic waves and relax the ion beams. Our study reveals a clear energy transfer channel from MHD-scale slow-mode waves to kinetic-scale ion acoustic waves.

\end{abstract}



\section{Introduction}
Plasmas support collective oscillations across multiple spatial scales, from large-scale magnetohydrodynamic (MHD) waves to kinetic-scale ion and electron plasma waves. Understanding how energy is transferred among these disparate spatial scales represents one of the most compelling challenges in space and astrophysical plasma physics. To investigate this cross-scale energy transfer in situ, modern spacecraft missions employ multi-point constellations, including the four-spacecraft Cluster mission \citep{Escoubet01}, the five-spacecraft THEMIS mission \citep{Angelopoulos08:ssr}, and the four-spacecraft Magnetospheric Multiscale (MMS) mission \citep{Burch16}. Cross-scale energy transfer events occur predominantly in plasma boundary regions---such as shocks, plasma jets, and current sheets/flux ropes---where steep gradients in magnetic field and plasma density are present. One example occurs when mesoscale plasma jets, generated by reconnection outflows in Earth's magnetotail, propagate into the dipole-dominated inner magnetosphere as dipolarization fronts \citep[e.g.,][]{Angelopoulos92,Runov09grl}. These fronts excite field line resonances at ion scales \citep{Chaston14,cheng2020kinetic}, which subsequently drive intense electron beams via nonlinear Landau resonance \citep{Artemyev15:jgr:KAW} and Debye-scale electrostatic turbulence \citep{Malaspina18,liu2022cross}. Ultimately, dipolarization fronts decelerate and deposit their energy into the ambient thermal plasma through this energy transfer spanning multiple spatial scales \citep{Khotyaintsev11,Ergun15,An21:kaw,Ukhorskiy22:NatSR}. {\blue{Similar energy transfer processes are likely to occur near coronal jets generated by magnetic reconnection on the Sun, which contribute to coronal heating and the acceleration of solar wind particles \citep{Masuda94, raouafi2016solar}, as well as in other stellar environments involving astrophysical jets associated with shocks \citep{blandford1998jets}.}}

As another example of plasma boundary dynamics, the interaction between the shocked solar wind and Earth's magnetic field generates large-amplitude surface waves at the magnetopause, driven by Kelvin-Helmholtz instabilities. Within these surface waves, observations from the four-spacecraft Cluster mission revealed ion-scale fast magnetosonic waves carrying sufficient energy to account for observed ion heating \citep{moore2016cross}. More recently, the four-spacecraft MMS mission has observed a rich variety of complex plasma phenomena at the magnetopause, including Alfv\'enic turbulence \citep{stawarz2016observations}, ion beams and plasma heating \citep{sorriso2019turbulence}, and large-amplitude electrostatic waves \citep{wilder2016observations}. These seemingly disparate phenomena are interconnected through Alfv\'en-acoustic energy channeling \citep[e.g.,][]{bierwage2015alfven}: magnetic pressure gradients associated with large-amplitude Alfv\'en waves drive compressive perturbations \citep{hollweg1971density} that accelerate ion beams via nonlinear Landau resonance \citep{medvedev1998asymptotic,araneda2008proton,matteini2010kinetics,gonzalez2021proton}. These accelerated ion beams subsequently excite Debye-scale ion acoustic waves, which relax the beams and produce parallel ion heating, thereby terminating the energy transfer chain \citep{valentini2008cross,valentini2014nonlinear,an2024cross}. This Alfv\'en-acoustic energy channeling mechanism may also operate around magnetic discontinuities in the solar wind \citep{mozer2020large,graham2021kinetic,woodham2021enhanced,malaspina2024frequency,vidal2025search} and, more broadly, in stellar wind environments.

In this Letter, we use observational data from the MMS mission to demonstrate that Debye-scale ion acoustic waves are excited at the recurring peaks of total magnetic field in obliquely propagating MHD-scale slow-mode waves, leading to preferential ion heating parallel to magnetic field lines. Through event-oriented hybrid-kinetic simulations, we show that bulk velocity perturbations associated with these large-amplitude slow-mode waves become comparable to ion thermal velocities at the magnetic field peaks. These large bulk velocity perturbations are manifested in velocity space as two distinct ion streams, which subsequently excite Debye-scale ion acoustic waves and produce preferential ion heating in the parallel direction. Our simulation results provide theoretical support for the interpretation of cross-scale energy transfer observed in the spacecraft data.

\section{Spacecraft observations}

This study utilizes observations from the MMS mission, which comprises four satellites arranged in a tetrahedral configuration \citep{Burch16}. Magnetic field measurements are provided by the fluxgate magnetometer (FGM) at 128 samples/s \citep{Russell16:mms} and the search-coil magnetometer (SCM) at 8192 samples/s in burst mode \citep{LeContel16}. Electric field measurements are obtained from a sensor suite comprising two axial and four spin-plane double-probe sensors (ADP and SDP) \citep{Ergun16:ssr, Lindqvist16}. Plasma measurements are acquired using the Fast Plasma Investigation (FPI) instrument suite \citep{Pollock16:mms}.

\begin{figure}[tphb]
\centering
\includegraphics[width=0.6\textwidth]{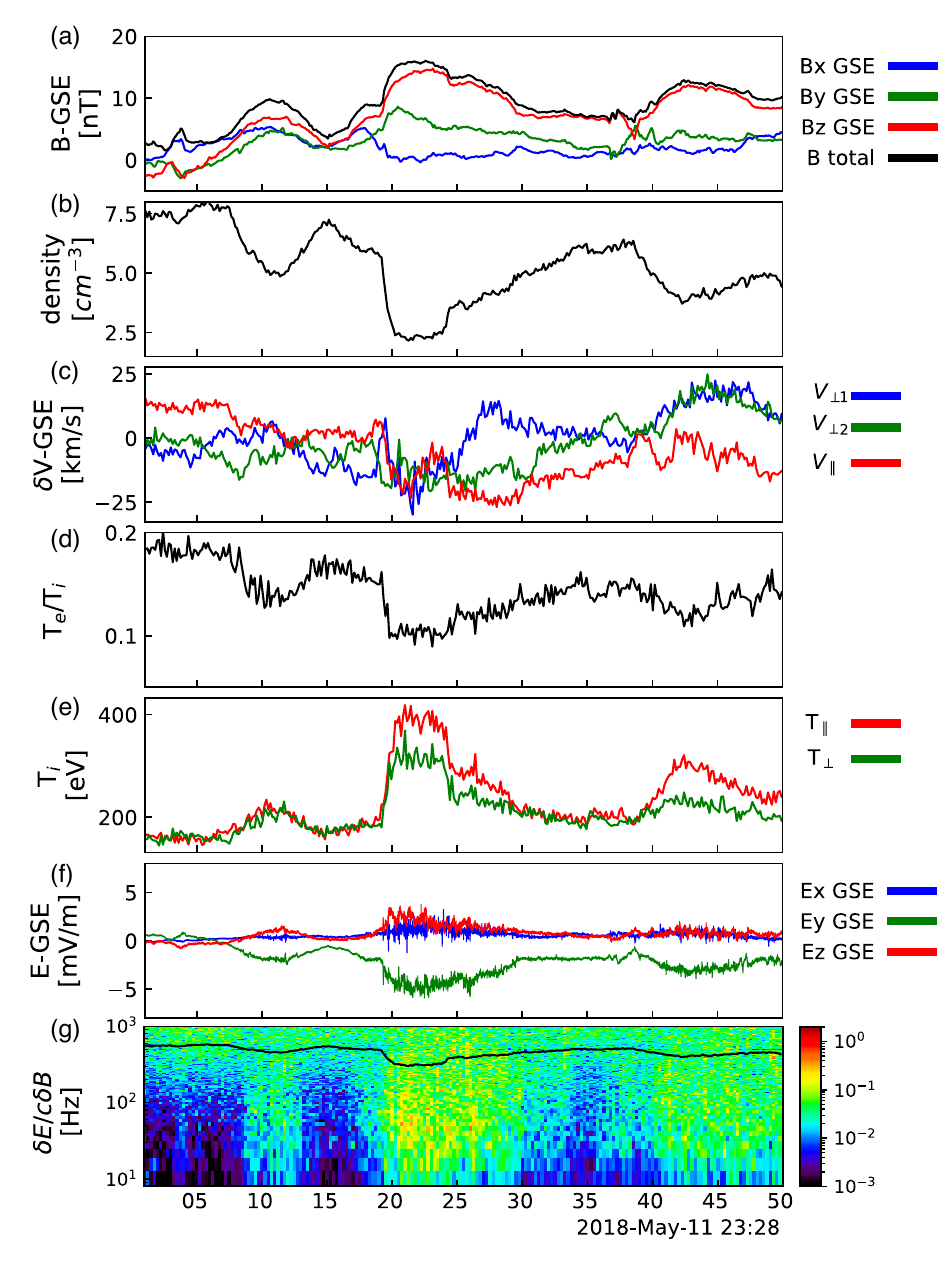}
\caption{\label{fig1}MMS observations of slow-mode MHD waves accompanied by electrostatic perturbations. (a) Total magnetic field magnitude (black line) and its three components in GSE coordinates. (b) Ion density. (c) Three components of plasma velocity perturbations with bulk velocity removed. $V_\parallel$ is the velocity component parallel to the background magnetic field. The perpendicular velocity is decomposed into two components, $V_{\perp1}$ and $V_{\perp2}$. Among them, $V_{\perp2}$ is oriented approximately along the propagation direction of the slow-mode wave. (d) Electron-to-ion temperature ratio. (e) Ion parallel and perpendicular temperatures. (f) Three components of electric field time series. (g) Electric field power normalized by magnetic field power. The black line shows the ion plasma frequency $f_{pi}$. (h) Ion parallel velocity distributions obtained from integrating phase space densities measured in the 3D velocity space (energy, pitch angle, gyrophase)}
\end{figure}

On 11 May 2018, the MMS satellites crossed Earth's magnetopause boundary around 23:29 UT, observing slow-mode MHD waves accompanied by electrostatic perturbations. Figure \ref{fig1} provides an overview of this event. Large-amplitude slow-mode waves were identified during 23:28–23:29 UT based on the anti-correlation between magnetic field strength and plasma density $n_i$ (the magnetic field strength is also anti-correlate with pressure)\citep{book:Kivelson&Russell95}. The observed slow-mode waves had a frequency of $f_{\mathrm{sc}} = 0.05$\,Hz in the spacecraft frame and a normalized amplitude of $\vert \delta B/B \vert \sim 0.5$, {\blue{where $\delta B$ is the wave amplitude}} and $B$ presents the background magnetic field calculated by averaging the total magnetic field over the event interval.

We applied the minimum variance analysis (MVA) method to determine the slow-mode wave propagation direction \citep{Sonnerup68,bookISSI:Sonnerup}. {\blue{The MVA method determines the principal directions of variance and their corresponding eigenvalues from single-satellite magnetic field measurements. The wave propagation direction is identified as the direction of minimum variance of the magnetic field perturbations \citep{Sonnerup68}.}} The waves propagated nearly perpendicular to the background magnetic field, with the angle between the wave vector $\vec{k}$ and the background magnetic field of $\theta_{kb} = 81.3^\circ$. Due to the ultra-low frequency of the waves, the observed spacecraft-frame frequency is dominated by Doppler-shift effects. The wavenumber can be calculated as $\vert k \vert = f_{\mathrm{sc}} / (V \cos\theta_{kv})\sim 0.05/d_i$, where $V=290$\,km/s is the plasma bulk velocity, $\theta_{kv}=76.1^\circ$ is the angle between the wave vector and the bulk velocity direction, {\blue{and  l.}}

Broadband electrostatic waves are also observed [Figures \ref{fig1}(f) and \ref{fig1}(g)], with frequencies ranging from $10$\,Hz up to the local ion plasma frequency ($f_{pi} \approx 500$\,Hz). These waves are identified as ion acoustic waves (IAWs), using the method described by \cite{Vasko22:grl,Wang20:apjl} (detailed analysis is provided in the Appendix A). The observed IAWs propagate at approximately $30^\circ$ to the background magnetic field, with a phase speed $v_{ph}$ in the plasma frame comparable to the local ion acoustic speed $V_c$, and wavelengths around $10\, \lambda_D$, where $\lambda_D$ is the local Debye length.

In contrast to the survival condition for IAWs in a Maxwellian plasma requiring an electron-to-ion temperature ratio $T_e / T_i \gg 1$, the observed IAWs survive in an environment with $T_e / T_i < 1$ [Figure \ref{fig1}(d)]. This observation indicates that the ion velocity distributions deviate significantly from Maxwellian distributions \citep[e.g.,][]{valentini2011short,an2024cross}, thereby reducing ion Landau damping of IAWs at their propagation velocities. Such non-Maxwellian ion velocity distributions were observed by MMS [see Figure \ref{fig1}(h)]. By comparison, electron Landau damping is always weak because the electron thermal velocity greatly exceeds IAW propagation velocities.

The intensity of these IAWs is enhanced near the peaks of the magnetic field strength associated with the slow-mode waves (around 23:28:25 UT and 23:28:45 UT). This correlation between IAW enhancement and slow-mode waves suggests that the IAW generation is associated with and possibly driven by the large-amplitude MHD-scale waves. The IAW excitation regions are characterized by enhanced ion temperatures (ion heating) accompanied by parallel temperature anisotropy [Figure \ref{fig1}(e)], as well as shear and compressional velocity perturbations in both parallel and perpendicular directions relative to the background magnetic field [Figure \ref{fig1}(c)]. Here, velocity perturbations are defined as the observed flow velocity with the average background bulk velocity removed.

\section{Kinetic simulations}

To investigate the generation of electrostatic waves and their relationship with slow-mode waves, we performed an event-oriented simulation using the Hybrid-VPIC code \citep{Le23}. This hybrid approach treats electrons as a massless fluid, while modeling ions as fully kinetic particles. The simulation domain has two dimensions in configuration space ($x \in [0, 250]\,d_i$, $y \in [0, 15]\,d_i$) and three dimensions in velocity space $(V_x, V_y, V_z)$. {\blue{We applied fully periodic boundary conditions to both fields and particles.}} We used a time step of $0.004\, \omega_{ci}^{-1}$, where $\omega_{ci}$ is the proton cyclotron frequency. {\blue{The simulation cell size is $\Delta x=0.12 d_i$ and $\Delta y=0.06 d_i$ with $400$ particles per cell. Hybrid codes inherently assume quasi-neutrality, which prevents them from resolving Debye-scale structures. Consequently, the hybrid approach cannot capture the wide spatial range from the slow-mode wave scale down to the Debye length. The energy accumulation near the grid scale is affected by numerical damping and does not represent the true kinetic physics at Debye scales. Within these limitations, our simulation can still capture longer-wavelength IAWs and demonstrate energy transfer from fluid scales to kinetic scales, providing insight into wave generation mechanisms at ion scales and above \citep[e.g.,][]{an2024cross}.}}

\begin{figure}[tphb]
\centering
\includegraphics[width=0.9\textwidth]{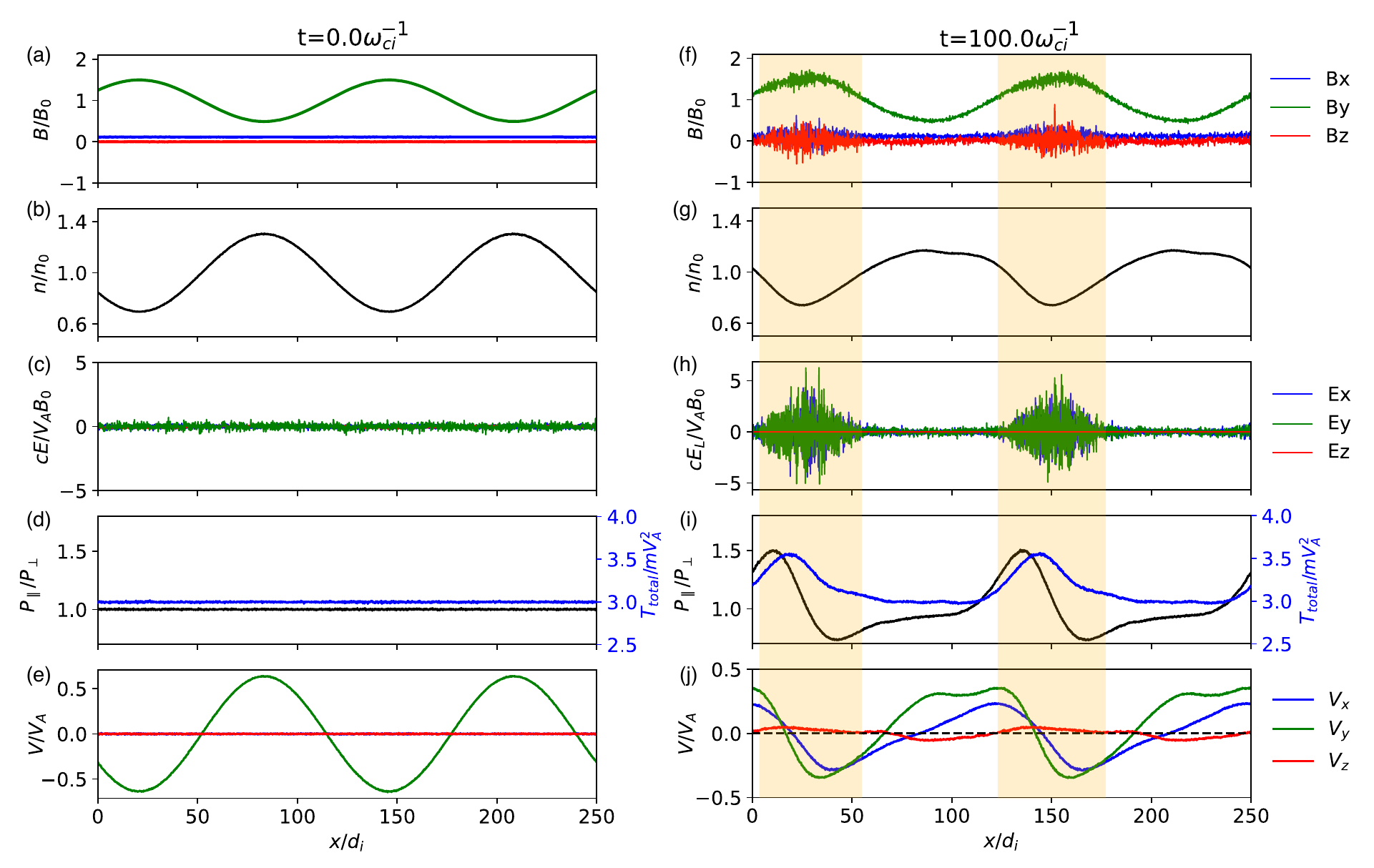}
\caption{\label{fig2}{\blue{Simulation results at two different times}}. Electromagnetic field and plasma properties at $t=0$ (left column) and $t=100\,\omega_{ci}^{-1}$ (right column) from the simulation. (a,f) Three components of magnetic field: blue, green, and red lines represent $B_x$, $B_y$, and $B_z$, respectively. (b,g) Density perturbation. (c,h) Three components of longitudinal electric field $\mathbf{E}_L$ with $\mathbf{k} \times \mathbf{E}_L = 0$. (d,i) Parallel-to-perpendicular pressure ratio and total temperature. (e,j) Three components of flow velocities.}
\end{figure}

From our event, the ratio between electron temperature ($T_e$) and ion temperature ($T_i$) is approximately $0.15$ on average, and the average ion beta is $\beta_i = n_i T_i / (B^2 / 8\pi) \approx 6$. This corresponds to a ratio of Alfv\'en speed to ion thermal speed of $V_A / V_{Ti} = \sqrt{2 / \beta} \approx 0.6$, while the ion acoustic speed relative to ion thermal speed is $V_c / V_{Ti} = \sqrt{(5/3) + T_e / T_i} \approx 1.3$. Based on these observations, we set the initial simulation conditions with $T_e/T_i = 0.15$ and $\beta_i = 6$. The complete initial field and plasma configuration at $t=0$ is shown in Figures \ref{fig2}(a)-\ref{fig2}(e). The initial pressure is anti-correlated with magnetic field strength with $P_\parallel/P_\perp=1$. The background magnetic field $B_0$ lies in the $x$–$y$ plane, oriented nearly along the $+y$ direction at an angle of $83.3^\circ$ with respect to the $+x$ axis. We initialize a slow-mode wave perturbation based on the observation data, propagating in the $+x$ direction nearly perpendicular to $B_0$. The magnetic field perturbation components are given by:
\begin{align}
    & \delta B_\parallel = \delta B \cos(4\pi x/L_x + \phi) , \\
    & \delta B_\perp = -\frac{k_\parallel}{k_\perp} \delta B \cos(4\pi x/L_x + \phi) \ll \delta B_{\parallel} ,
\end{align}
where $\delta B_{\parallel}$ and $\delta B_{\perp}$ are the wave perturbations parallel and perpendicular to the background magnetic field in the $x$-$y$ plane, respectively. The wave amplitude is $\delta B = 0.5\,B_0$, the wavenumber is $k = 4\pi/L_x = 0.05d_i$, and the randomly selected initial phase is $\phi = 2\pi/3$.
The corresponding initial density perturbation is $\delta n = -n_0 (\delta B/B_0) V_A^2 / V_c^2$. Ions are initialized with a Maxwellian distribution and include a velocity perturbation from the slow-mode wave: $\delta V_\parallel = -(\delta B_\parallel/B_0) \sqrt{V_c^2 + V_A^2} V_A / V_c$.
All simulation quantities are normalized as follows: magnetic field to $B_0$, time to $\omega_{ci}^{-1}$, velocity to $V_A$, density to $n_0$, and spatial scale to $d_i$. The electric field is consequently normalized to $V_A B_0 / c$, where $c$ is the speed of light.

Electric field perturbations develop self-consistently as the simulation evolves. By $t = 100\,\omega_{ci}^{-1}$, large-amplitude, small-scale electric field perturbations emerge near the magnetic field peaks at $x \approx 20\,d_i$ and $x \approx 140\,d_i$. Small-scale magnetic field perturbations appear at these same locations, indicating the presence of small-scale electromagnetic waves. To separate the electromagnetic and electrostatic components, we decomposed the electric field into longitudinal and transverse parts with respect to the wavevector $\vec{k}$ in the $x$–$y$ plane. We calculated the electrostatic (longitudinal) component in wavenumber space using $\vec{E}_L=(\vec{k}\cdot \vec{E})\vec{k}/k^2$ and transformed the result back to real space. The resulting electrostatic component [Figure \ref{fig2}(h)] shows perturbations that are maximized near the magnetic field peaks, with comparable amplitudes in both $x$ and $y$ directions. {\blue{Importantly, ion heating ($T_{\mathrm{total}}$) develops in the same regions where electrostatic waves are excited, successfully reproducing key features observed by MMS [Figures \ref{fig1}(a), \ref{fig1}(e), \ref{fig1}(f), and \ref{fig1}(g)]. The parallel pressure anisotropy ($P_\parallel/P_\perp > 1$) does not fully capture observations at $t = 100\omega_{ci}^{-1}$, as the anisotropy falls below unity near the edge of the electrostatic wave region. However, at later times, the parallel pressure anisotropy becomes consistent with the observed spatial distribution, remaining greater than one throughout the wave fluctuation region (Appendix B).}}

\begin{figure}[h!]
\centering
\includegraphics[width=0.6\textwidth]{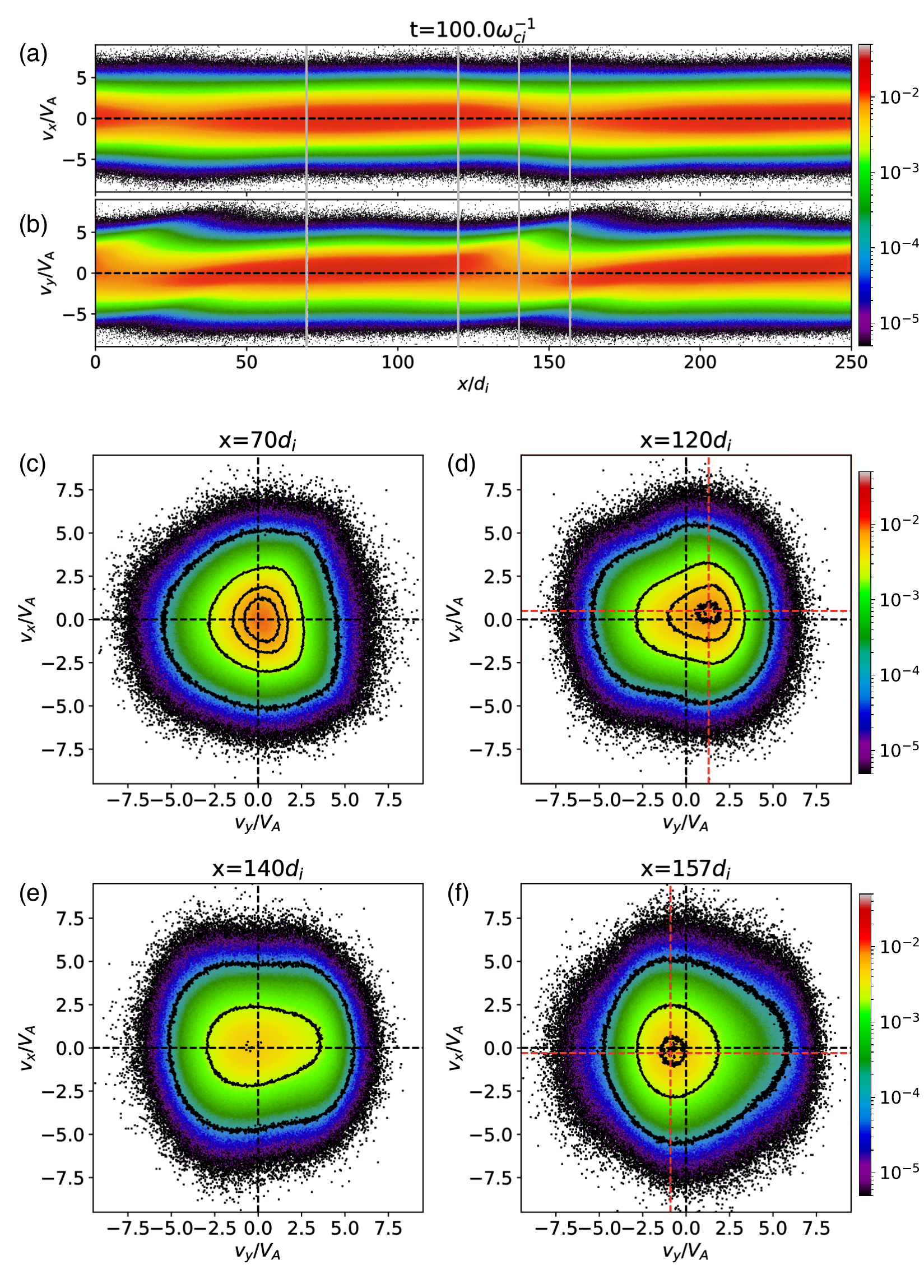}
\caption{\label{fig3}{\blue{Simulation results of ion phase density and velocity distributions at $t=100\,\omega_{ci}^{-1}$.}} (a, b) Ion phase space density; (c-f) Velocity distribution in $v_x-v_y$ plane (integrated over $v_z$), at $x=70 d_i$, $x=120d_i$, $x=140d_i$, $x=157d_i$, respectively. The {\blue{red}} dashed lines in panel (d, f) indicate the location of ion distribution peaks. }
\end{figure}

The locations for electrostatic wave excitation are intimately linked to flow velocity gradients. Initially, consistent with the slow-mode wave dispersion relation, the ion drift velocity component, $V_x$, is nearly zero while $V_y$ remains $180^\circ$ out of phase with the large-amplitude magnetic field perturbations [Figure \ref{fig2}(e)]. As the system evolves, the $V_x$ perturbation amplitude grows due to the magnetic pressure gradient, and the magnetic tension force drives the steepening of $V_y$ near the magnetic field peaks at $x \approx 20 d_i$ and $x \approx 140 d_i$. By $t = 100 \omega_{ci}^{-1}$, both $V_x$ and $V_y$ develop large spatial gradients near these magnetic field peaks, exhibiting compressions ($\partial_x V_x \neq 0$) and flow shears ($\partial_x V_y \neq 0$). This behavior is exemplified around $x = 140 d_i$, where the ion drift velocities $V_x$ and $V_y$ are positive for $x < 140 d_i$, but become negative for $x > 140 d_i$. This velocity reversal indicates counter-streaming ion flows on the two sides of the magnetic-field peak, a feature that closely matches the flow patterns observed by MMS [Figure \ref{fig1}(c)].

To understand kinetic picture underlying the behavior of flow velocities, we examine the ion velocity distributions on $v_x$-$v_y$ plane. Figure \ref{fig3} presents the ion phase-space density and velocity distributions at four different $x$ locations at $t = 100 \omega_{ci}^{-1}$. On either side of $x \approx 140 d_i$, ions exhibit core distributions with velocities of opposite signs [Figures \ref{fig3}(d) and \ref{fig3}(f)]. At $x = 120 d_i$, the ion velocity distribution peaks at $v_x \approx 0.5 V_A$ and $v_y \approx 1.3 V_A$, whereas at $x = 160 d_i$, the maximum occurs at $v_x \approx -0.3 V_A$ and $v_y \approx -V_A$. These distributions confirm counter-streaming drift toward $x = 140 d_i$ in both $x$ and $y$ directions. Such flows transport ions from high-density toward low-density regions, thereby reducing the density perturbation amplitude compared to the initial stage of the simulation. At the convergence point $x = 140 d_i$, the opposing flows yield a near-zero mean velocity and create a plateau-like velocity distribution [Figure \ref{fig3}(e)]. Here, the electrostatic perturbations reach their maximum and ion heating is evident. Similar behavior occurs around $x = 20 d_i$. In contrast, near $x = 70 d_i$, where the magnetic field reaches its minimum, the mean velocity is zero without counter-streaming flows. Instead, ions display a Maxwellian-type distribution centered near zero velocity, and electrostatic perturbations are nearly absent.

The simulation results indicate that counter-streaming ion flows in both the $x$ and $y$ directions are likely responsible for the excitation of IAWs. Classical excitation mechanisms for IAWs include the current-driven instability and the ion–ion acoustic instability \citep{buneman1959dissipation,Davidson70,Gary87JPP}. The current-driven instability requires a significant drift between ions and electrons. In this event, with $T_i/T_e \sim 6$, triggering the current-driven instability would require an ion-to-electron drift comparable to the electron thermal velocity ($\sim 40 V_A$), which is physically unlikely under these conditions. The ion–ion acoustic instability, on the other hand, requires two ion populations streaming toward each other, typically a dense cold core and a hot beam. Although a parallel ion beam is observed in this event, such a parallel beam cannot account for IAWs perturbations of comparable amplitude in directions perpendicular to magnetic field lines. Our simulation results suggest that the observed plateau-like velocity distributions [Figure \ref{fig3}(e)] and counter-streaming flows in both $x$ and $y$ directions [Figures \ref{fig3}(d) and \ref{fig3}(f)]---distinct from the classic core–beam configuration---may represent a complementary mechanism for IAW generation.

{\blue{Conventionally, IAWs are thought to be generated under conditions of a large electron-to-ion temperature ratio. When $T_e/T_i$ is small, strong ion Landau damping typically suppress wave growth. However, in our case, despite $T_e/T_i < 1$ and the phase velocity of the electrostatic turbulence ($V_{ph}$ being comparable to the ion thermal speed ($V_{Ti}$), a plateau forms near the ion thermal speed. This plateau reduces the phase space density gradient and thereby weakens ion Landau damping, enabling the generation of ion-bulk waves (IAWs propagating at the ion thermal speed). These conditions resemble those discussed by \cite{valentini2011short} and \cite{valentini2014nonlinear}.}}

{\blue{Although the simulation results demonstrate a possible mechanism for IAW generation through counter-streaming ion drifts in both parallel and perpendicular directions relative to the background magnetic field, evidently the existing theories do not adequately describe the details of this process. The simulation also reveals additional complexities, particularly the emergence of small-scale electromagnetic perturbations near the magnetic field peaks [Figure \ref{fig2}(f)]. Resolving these questions will require developing a comprehensive linear kinetic theory that incorporates ion drifts in both parallel and perpendicular directions driven by large-amplitude slow-mode waves. However, such theoretical development presents significant challenges: ion perpendicular drifts arising from spatial gradients of density and magnetic field must be self-consistently incorporated as background flows, with linear instability analysis then performed within this modified equilibrium---an approach analogous to certain aspects of the linear kinetic analysis of lower-hybrid drift instability \citep{huba1976effects,Huba77}. This multi-step theoretical framework is beyond the current scope of this work.}}

\section{Summary}

\begin{figure}[tphb]
    \centering
    \includegraphics[width=0.8\linewidth]{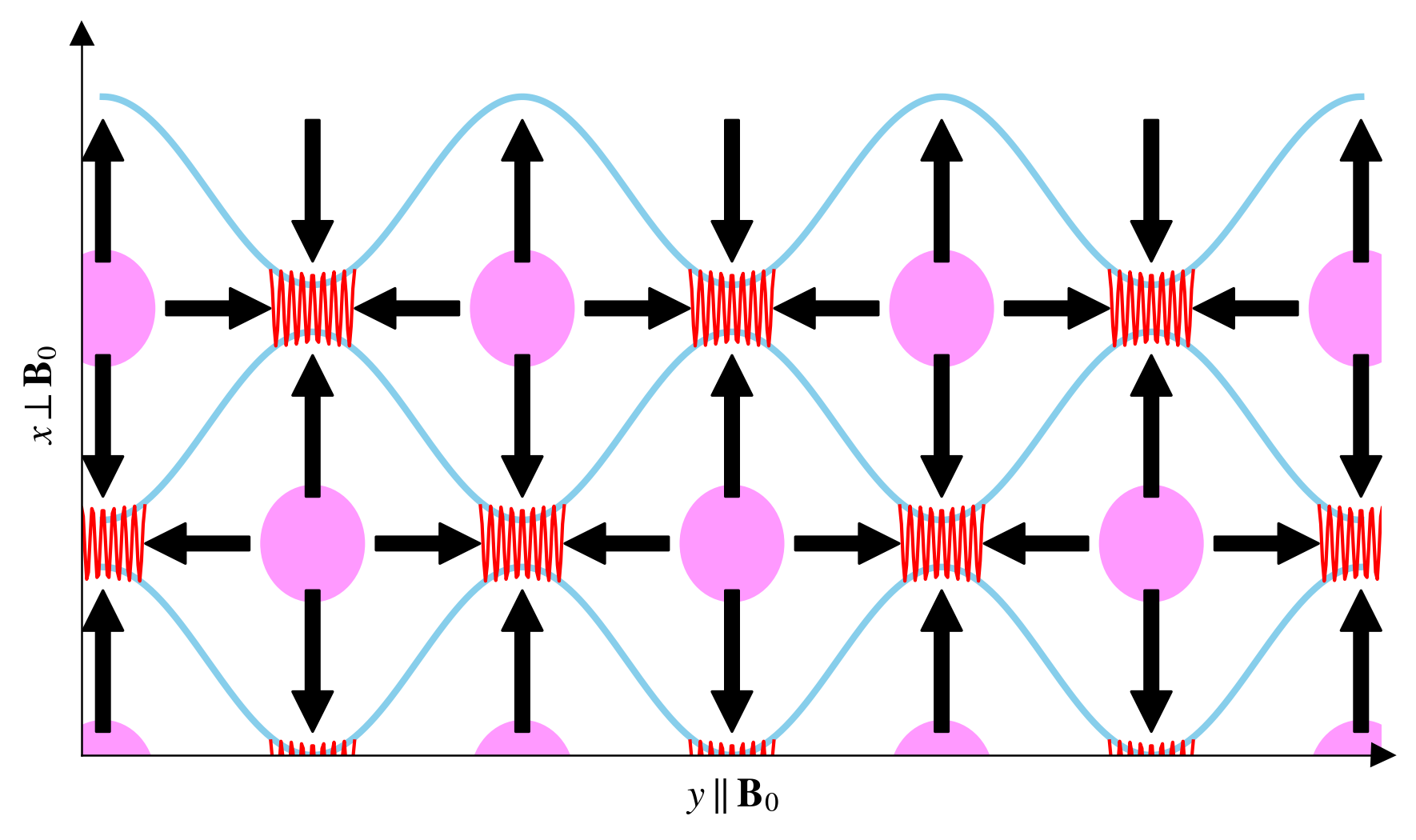}
    \caption{Schematic illustration of magnetic field and ion flow perturbations in MHD-scale slow-mode waves and their coupling to kinetic-scale IAWs. Blue lines represent slow-mode magnetic field perturbations superimposed on a background magnetic field $\mathbf{B}_0$, forming MHD-scale magnetic mirrors. The aspect ratio is not to scale for visualization clarity: the spatial scale parallel to $\mathbf{B}_0$ is much longer than the perpendicular scale, i.e., $k_x \gg k_y$. Magenta ellipses indicate higher density regions at magnetic mirror centers. Counter-streaming shear flows parallel to $\mathbf{B}_0$ ($\partial_x V_y \neq 0$) and compressional flows perpendicular to $\mathbf{B}_0$ ($\partial_x V_x \neq 0$) converge toward magnetic field peaks, exciting kinetic-scale IAWs (red sinusoidal curves).}
    \label{fig:sketch}
\end{figure}

In summary, this study presents direct observational and computational evidence for energy transfer from MHD-scale slow-mode waves to kinetic-scale IAWs. At the magnetopause boundary, enhanced IAW activity and non-adiabatic ion heating in velocity space correlate spatially with magnetic field peaks associated with slow-mode waves, suggesting that the field and particle energy at the kinetic scale originates from MHD-scale magnetic field fluctuations. Using parameters derived from these observations, we performed hybrid-kinetic simulations that successfully reproduced the observed counter-streaming ion flows both parallel and perpendicular to magnetic field lines and the formation of velocity distribution plateaus. The spacecraft observations and accompanied simulation reveal that ion drifts driven by slow-mode waves create the free energy necessary for IAW excitation through kinetic instabilities. Our results demonstrate a cross-scale energy transfer process from fluid to kinetic scales, in which energy is channeled into the ion population, leading to substantial ion heating. Figure \ref{fig:sketch} provides a schematic illustration of these findings and the physical configuration governing this cross-scale energy transfer process.

More broadly, given the prevalence of slow-mode structures and turbulence in the solar wind \citep{roberts2006slow,howes2012slow}, the cross-scale energy transfer mechanism associated with slow-mode waves provides an effective pathway for converting {\blue{bulk kinetic energy}} into ion thermal energy near general plasma boundaries throughout the heliosphere. Furthermore, because large-amplitude, Debye-scale IAWs scatter thermal electrons in velocity space \citep{kamaletdinov2024nonlinear,zanelli2025flat}, this cross-scale mechanism may facilitate momentum exchange between ions and electrons in current sheets, thereby enabling collisionless current dissipation and magnetic reconnection \citep{coppi1971processes,coroniti1977magnetic,sagdeev19791976,smith1972current}.


\begin{acknowledgments}
This work was supported by NASA grants NO.~80NSSC22K1634 and NSF grant NO.~2108582. We acknowledge MMS data (including FGM, EDP, and FPI) obtained from \url{https://lasp.colorado.edu/mms/sdc/public/}. Data access and processing was done using SPEDAS V4.1 \citep{Angelopoulos19}. We would like to acknowledge high-performance computing support from Derecho (\url{https://doi.org/10.5065/qx9a-pg09}) provided by NCAR's Computational and Information Systems Laboratory, sponsored by the National Science Foundation \citep{derecho}. We wish to thank Anton Artemyev and Marco Velli for useful discussions.
\end{acknowledgments}

%






\appendix

\section{Interferometry technique}

\begin{figure}[h!]
\centering
\includegraphics[width=0.6\textwidth]{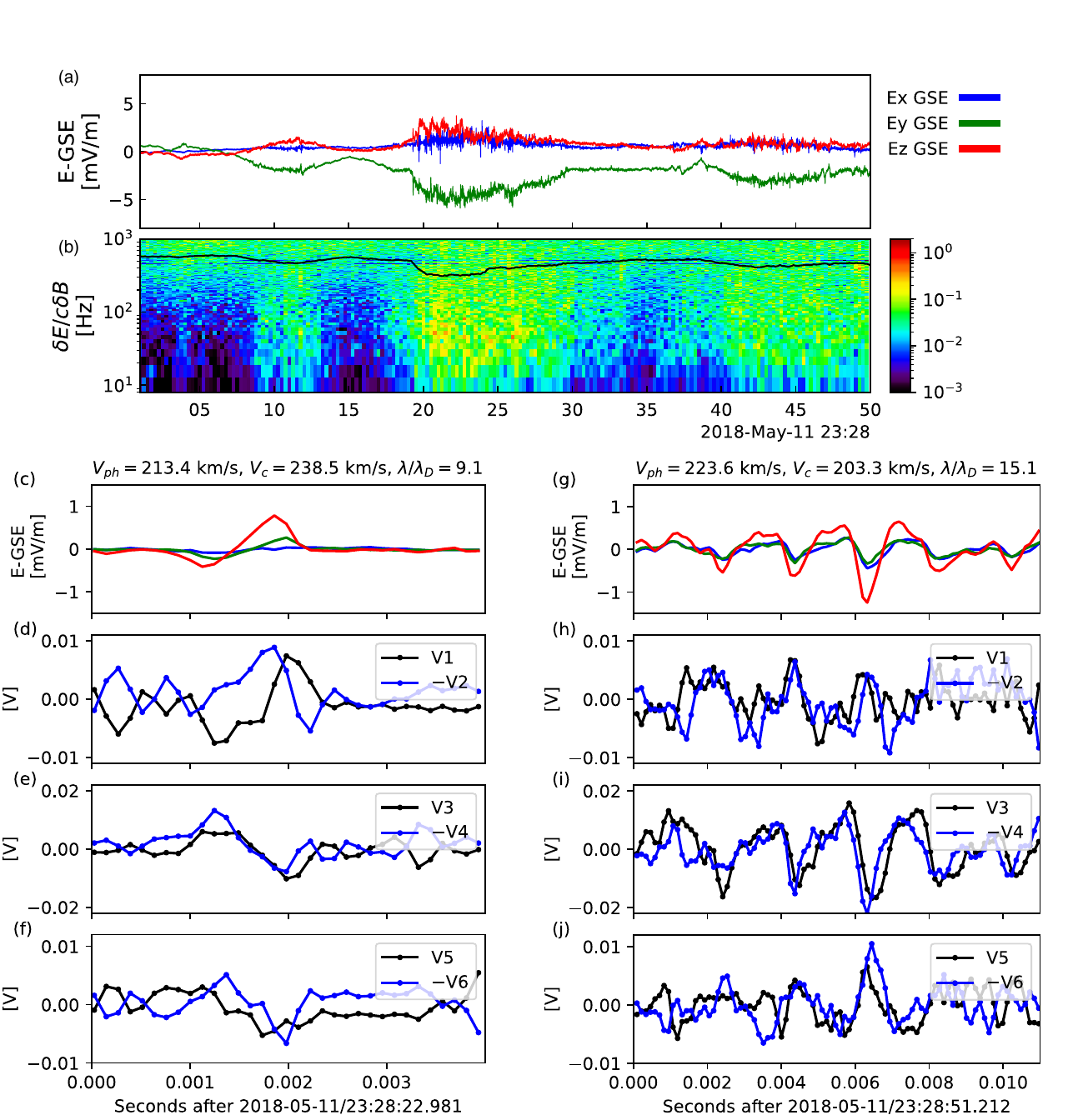}
\caption{\label{figS1} (a) Three components of electric field time series. (b) Electric field power normalized by magnetic field power. The black line indicate the ion plasma frequency. (c-j) The electric field and associated voltage signals in three pairs of opposing voltage-sensitive
probes. The voltage signals $V_1$ vs. -$V_2$ and $V_3$ vs -$V_4$ in the two perpendicular directions in the spin plane. The voltage signals $V_5$ and -$V_6$ along the spin axis.}
\end{figure}

The propagation characteristics of ion acoustic waves (IAWs) are investigated using the interferometry technique. {\blue{Figure \ref{figS1} shows two examples of the three pairs of voltage measurements (panels d–f and panels h–i). The voltage signals of $V_1$ vs.~$-V_2$, $V_3$ vs.~$-V_4$, and $V_5$ vs.~$-V_6$ correspond to three orthogonal directions. The physical separations between the voltage sensors in the spin plane are $l_{12} = l_{34} = 60$ m, and the distance along the spin axis is $l_{56} = 29.2$ m. The three components of the electric field along the boom directions are calculated as $E_{ij} = (V_i - V_j) / (2l_{ij})$, where $V_i$ and $V_j$ are the voltages measured by opposing probes separated by a distance $l_{ij}$. The propagation velocity and direction of the IAWs can be determined from the time delays $\Delta t_{ij}$ between the voltage signals measured by the opposing probes \cite[e.g.,][]{Vasko22:grl}.}}:

\begin{eqnarray}
    \frac{1}{V_s^{2}} = \frac{\Delta t_{12}^2}{l_{12}^2} + \frac{\Delta t_{34}^2}{l_{34}^2} + \frac{\Delta t_{56}^2}{l_{56}^2} & , \\
    \vec{\mathbf{k}} = \left( -\frac{V_s \Delta t_{12}}{l_{12}}, -\frac{V_s \Delta t_{34}}{l_{34}}, -\frac{V_s \Delta t_{56}}{l_{56}} \right) & .
\end{eqnarray}

{\blue{where $V_s$ is the wave propagation velocity in the spacecraft frame, and $\vec{\mathbf{k}}$ represents the propagation direction of the IAWs. In the two examples shown in Figure \ref{figS1}, the time delay was determined by identifying the time shift that yields the highest correlation between the two voltage signals. Using the local ion flow velocity measured by MMS, we calculated the IAW phase velocity in the plasma frame as $V_{ph} = V_s - \vec{k} \cdot \vec{V}$. The phase velocity in the plasma frame and the wavelength normalized to the Debye length ($\lambda / \lambda_D$) are shown in Figure \ref{figS1}. The propagation speed of the IAWs is comparable to the ion acoustic speed ($V_c$), which satisfies the relation $V_c = \sqrt{(5/3 + T_e/T_i)} \cdot V_{Ti}$, where $V_{Ti}$ is the ion thermal velocity. The propagation angle is approximately $30^\circ$ relative to the background magnetic field.} }

\section{Simulation results at later times}

\begin{figure}[h]
\centering
\includegraphics[width=0.85\textwidth]{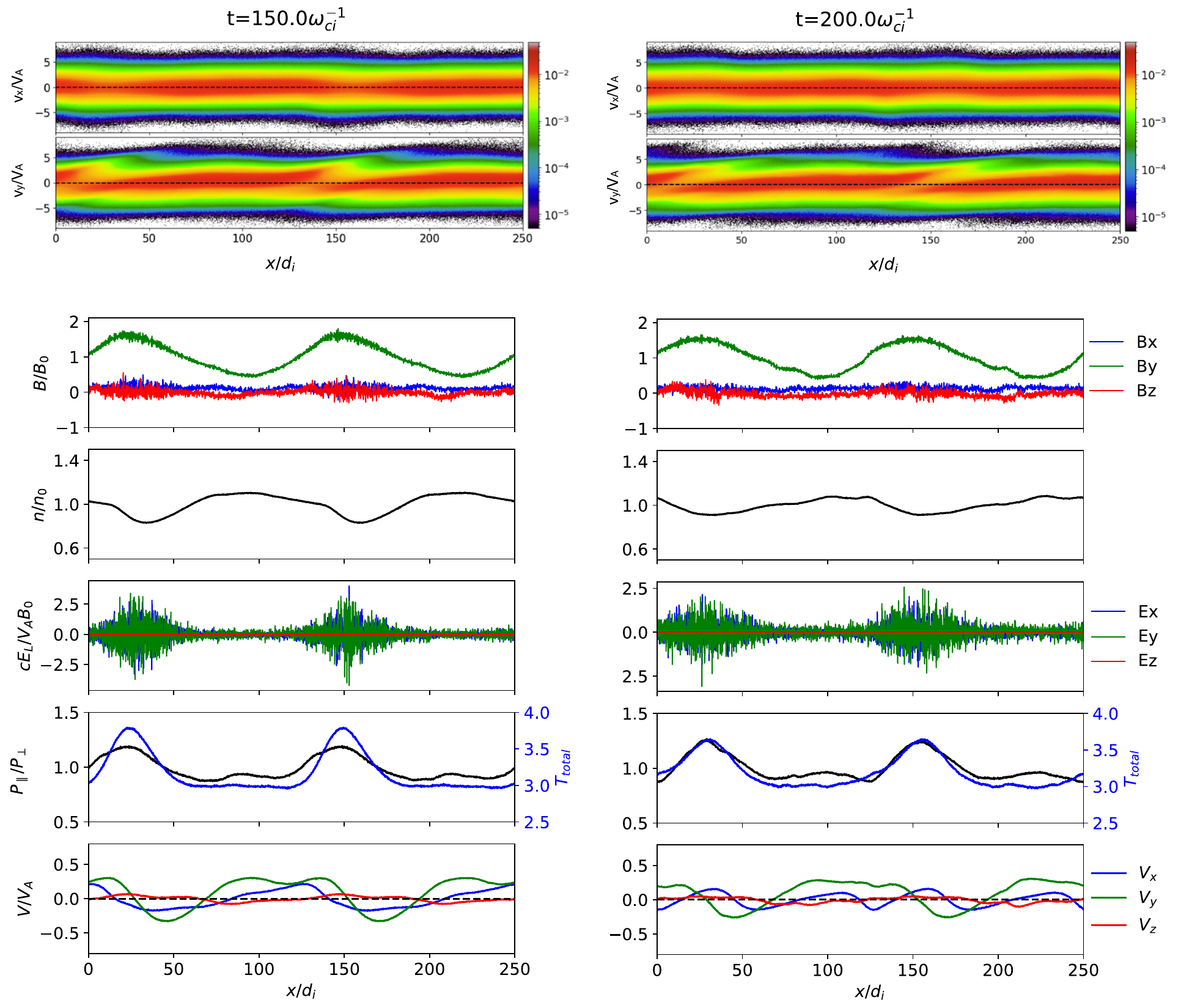}
\caption{\label{Fig5} {\blue{Simulation results of ion phase density, electromagnetic field and plasma properties at $t=150\,\omega_{ci}^{-1}$ (left) and $t=200\,\omega_{ci}^{-1}$ (right). Similar format as Figure 2 and Figure 3}}}
\end{figure}

{\blue{Figure \ref{Fig5} shows the simulation results of ion phase density, electromagnetic field and plasma properties at $150 \omega_{ci}^{-1}$ and $200 \omega_{ci}^{-1}$.}}

\begin{thebibliography}{}
\expandafter\ifx\csname natexlab\endcsname\relax\def\natexlab#1{#1}\fi
\providecommand{\url}[1]{\href{#1}{#1}}
\providecommand{\dodoi}[1]{doi:~\href{http://doi.org/#1}{\nolinkurl{#1}}}
\providecommand{\doeprint}[1]{\href{http://ascl.net/#1}{\nolinkurl{http://ascl.net/#1}}}
\providecommand{\doarXiv}[1]{\href{https://arxiv.org/abs/#1}{\nolinkurl{https://arxiv.org/abs/#1}}}

\bibitem[{{An} {et~al.}(2024){An}, {Artemyev}, {Angelopoulos}, {Liu}, {Vasko}, \& {Malaspina}}]{an2024cross}
{An}, X., {Artemyev}, A., {Angelopoulos}, V., {et~al.} 2024, \prl, 133, 225201, \dodoi{10.1103/PhysRevLett.133.225201}

\bibitem[{{An} {et~al.}(2021){An}, {Bortnik}, \& {Zhang}}]{An21:kaw}
{An}, X., {Bortnik}, J., \& {Zhang}, X.-J. 2021, Journal of Geophysical Research (Space Physics), 126, e28643, \dodoi{10.1029/2020JA028643}

\bibitem[{{Angelopoulos}(2008)}]{Angelopoulos08:ssr}
{Angelopoulos}, V. 2008, \ssr, 141, 5, \dodoi{10.1007/s11214-008-9336-1}

\bibitem[{{Angelopoulos} {et~al.}(1992){Angelopoulos}, {Baumjohann}, {Kennel}, {Coronti}, {Kivelson}, {Pellat}, {Walker}, {Luehr}, \& {Paschmann}}]{Angelopoulos92}
{Angelopoulos}, V., {Baumjohann}, W., {Kennel}, C.~F., {et~al.} 1992, \jgr, 97, 4027, \dodoi{10.1029/91JA02701}

\bibitem[{{Angelopoulos} {et~al.}(2019){Angelopoulos}, {Cruce}, {Drozdov}, {Grimes}, {Hatzigeorgiu}, {King}, {Larson}, {Lewis}, {McTiernan}, {Roberts}, {Russell}, {Hori}, {Kasahara}, {Kumamoto}, {Matsuoka}, {Miyashita}, {Miyoshi}, {Shinohara}, {Teramoto}, {Faden}, {Halford}, {McCarthy}, {Millan}, {Sample}, {Smith}, {Woodger}, {Masson}, {Narock}, {Asamura}, {Chang}, {Chiang}, {Kazama}, {Keika}, {Matsuda}, {Segawa}, {Seki}, {Shoji}, {Tam}, {Umemura}, {Wang}, {Wang}, {Redmon}, {Rodriguez}, {Singer}, {Vandegriff}, {Abe}, {Nose}, {Shinbori}, {Tanaka}, {UeNo}, {Andersson}, {Dunn}, {Fowler}, {Halekas}, {Hara}, {Harada}, {Lee}, {Lillis}, {Mitchell}, {Argall}, {Bromund}, {Burch}, {Cohen}, {Galloy}, {Giles}, {Jaynes}, {Le Contel}, {Oka}, {Phan}, {Walsh}, {Westlake}, {Wilder}, {Bale}, {Livi}, {Pulupa}, {Whittlesey}, {DeWolfe}, {Harter}, {Lucas}, {Auster}, {Bonnell}, {Cully}, {Donovan}, {Ergun}, {Frey}, {Jackel}, {Keiling}, {Korth}, {McFadden}, {Nishimura}, {Plaschke}, {Robert}, {Turner}, {Weygand}, {Candey}, {Johnson},
  {Kovalick}, {Liu}, {McGuire}, {Breneman}, {Kersten}, \& {Schroeder}}]{Angelopoulos19}
{Angelopoulos}, V., {Cruce}, P., {Drozdov}, A., {et~al.} 2019, \ssr, 215, 9, \dodoi{10.1007/s11214-018-0576-4}

\bibitem[{Araneda {et~al.}(2008)Araneda, Marsch, Adolfo, {et~al.}}]{araneda2008proton}
Araneda, J.~A., Marsch, E., Adolfo, F., {et~al.} 2008, Physical review letters, 100, 125003

\bibitem[{{Artemyev} {et~al.}(2015){Artemyev}, {Rankin}, \& {Blanco}}]{Artemyev15:jgr:KAW}
{Artemyev}, A.~V., {Rankin}, R., \& {Blanco}, M. 2015, \jgr, 120, 10, \dodoi{10.1002/2015JA021781}

\bibitem[{{Bierwage} {et~al.}(2015){Bierwage}, {Aiba}, \& {Shinohara}}]{bierwage2015alfven}
{Bierwage}, A., {Aiba}, N., \& {Shinohara}, K. 2015, \prl, 114, 015002, \dodoi{10.1103/PhysRevLett.114.015002}

\bibitem[{{Blandford}(1988)}]{blandford1998jets}
{Blandford}, R.~D. 1988, in Perspectives in Fluid Mechanics, ed. D.~{Coles}, Vol. 320, 14--30, \dodoi{10.1007/BFb0021115}

\bibitem[{Buneman(1959)}]{buneman1959dissipation}
Buneman, O. 1959, Physical Review, 115, 503

\bibitem[{{Burch} {et~al.}(2016){Burch}, {Moore}, {Torbert}, \& {Giles}}]{Burch16}
{Burch}, J.~L., {Moore}, T.~E., {Torbert}, R.~B., \& {Giles}, B.~L. 2016, \ssr, 199, 5, \dodoi{10.1007/s11214-015-0164-9}

\bibitem[{{Chaston} {et~al.}(2014){Chaston}, {Bonnell}, {Wygant}, {Mozer}, {Bale}, {Kersten}, {Breneman}, {Kletzing}, {Kurth}, {Hospodarsky}, {Smith}, \& {MacDonald}}]{Chaston14}
{Chaston}, C.~C., {Bonnell}, J.~W., {Wygant}, J.~R., {et~al.} 2014, \grl, 41, 209, \dodoi{10.1002/2013GL058507}

\bibitem[{Cheng {et~al.}(2020)Cheng, Lin, Perez, Johnson, \& Wang}]{cheng2020kinetic}
Cheng, L., Lin, Y., Perez, J., Johnson, J.~R., \& Wang, X. 2020, Journal of Geophysical Research: Space Physics, 125, e2019JA027062

\bibitem[{{Computational and Information Systems Laboratory}(2024)}]{derecho}
{Computational and Information Systems Laboratory}. 2024, Derecho: {HPE} {C}ray {EX} {S}ystem, Boulder, CO: National Center for Atmospheric Research.
\newblock \url{https://doi.org/10.5065/qx9a-pg09}

\bibitem[{Coppi \& Friedland(1971)}]{coppi1971processes}
Coppi, B., \& Friedland, A.~B. 1971, Astrophysical Journal, vol. 169, p. 379, 169, 379

\bibitem[{Coroniti \& Eviatar(1977)}]{coroniti1977magnetic}
Coroniti, F., \& Eviatar, A. 1977, Astrophysical Journal Supplement Series, vol. 33, Feb. 1977, p. 189-210., 33, 189

\bibitem[{{Davidson} {et~al.}(1970){Davidson}, {Krall}, {Papadopoulos}, \& {Shanny}}]{Davidson70}
{Davidson}, R.~C., {Krall}, N.~A., {Papadopoulos}, K., \& {Shanny}, R. 1970, \prl, 24, 579, \dodoi{10.1103/PhysRevLett.24.579}

\bibitem[{{Ergun} {et~al.}(2015){Ergun}, {Goodrich}, {Stawarz}, {Andersson}, \& {Angelopoulos}}]{Ergun15}
{Ergun}, R.~E., {Goodrich}, K.~A., {Stawarz}, J.~E., {Andersson}, L., \& {Angelopoulos}, V. 2015, \jgr, 120, 1832, \dodoi{10.1002/2014JA020165}

\bibitem[{{Ergun} {et~al.}(2016){Ergun}, {Tucker}, {Westfall}, {Goodrich}, {Malaspina}, {Summers}, {Wallace}, {Karlsson}, {Mack}, {Brennan}, {Pyke}, {Withnell}, {Torbert}, {Macri}, {Rau}, {Dors}, {Needell}, {Lindqvist}, {Olsson}, \& {Cully}}]{Ergun16:ssr}
{Ergun}, R.~E., {Tucker}, S., {Westfall}, J., {et~al.} 2016, \ssr, 199, 167, \dodoi{10.1007/s11214-014-0115-x}

\bibitem[{{Escoubet} {et~al.}(2001){Escoubet}, {Fehringer}, \& {Goldstein}}]{Escoubet01}
{Escoubet}, C.~P., {Fehringer}, M., \& {Goldstein}, M. 2001, Annales Geophysicae, 19, 1197, \dodoi{10.5194/angeo-19-1197-2001}

\bibitem[{{Gary} \& {Omidi}(1987)}]{Gary87JPP}
{Gary}, S.~P., \& {Omidi}, N. 1987, Journal of Plasma Physics, 37, 45, \dodoi{10.1017/S0022377800011983}

\bibitem[{Gonz{\'a}lez {et~al.}(2021)Gonz{\'a}lez, Tenerani, Matteini, Hellinger, \& Velli}]{gonzalez2021proton}
Gonz{\'a}lez, C., Tenerani, A., Matteini, L., Hellinger, P., \& Velli, M. 2021, The Astrophysical Journal Letters, 914, L36

\bibitem[{Graham {et~al.}(2021)Graham, Khotyaintsev, Vaivads, Edberg, Eriksson, Johansson, Sorriso-Valvo, Maksimovic, Sou{\v{c}}ek, P{\'\i}{\v{s}}a, {et~al.}}]{graham2021kinetic}
Graham, D.~B., Khotyaintsev, Y.~V., Vaivads, A., {et~al.} 2021, Astronomy \& Astrophysics, 656, A23

\bibitem[{Hollweg(1971)}]{hollweg1971density}
Hollweg, J.~V. 1971, Journal of Geophysical Research, 76, 5155

\bibitem[{Howes {et~al.}(2012)Howes, Bale, Klein, Chen, Salem, \& TenBarge}]{howes2012slow}
Howes, G., Bale, S., Klein, K., {et~al.} 2012, The Astrophysical Journal Letters, 753, L19

\bibitem[{Huba \& Wu(1976)}]{huba1976effects}
Huba, J., \& Wu, C. 1976, The Physics of Fluids, 19, 988

\bibitem[{{Huba} {et~al.}(1977){Huba}, {Gladd}, \& {Papadopoulos}}]{Huba77}
{Huba}, J.~D., {Gladd}, N.~T., \& {Papadopoulos}, K. 1977, \grl, 4, 125, \dodoi{10.1029/GL004i003p00125}

\bibitem[{Kamaletdinov {et~al.}(2024)Kamaletdinov, Vasko, \& Artemyev}]{kamaletdinov2024nonlinear}
Kamaletdinov, S.~R., Vasko, I.~Y., \& Artemyev, A.~V. 2024, Journal of Plasma Physics, 90, 905900201

\bibitem[{{Khotyaintsev} {et~al.}(2011){Khotyaintsev}, {Cully}, {Vaivads}, {Andr{\'e}}, \& {Owen}}]{Khotyaintsev11}
{Khotyaintsev}, Y.~V., {Cully}, C.~M., {Vaivads}, A., {Andr{\'e}}, M., \& {Owen}, C.~J. 2011, Physical Review Letters, 106, 165001, \dodoi{10.1103/PhysRevLett.106.165001}

\bibitem[{{Kivelson} \& {Russell}(1995)}]{book:Kivelson&Russell95}
{Kivelson}, M.~G., \& {Russell}, C.~T. 1995, {Introduction to Space Physics}, 586

\bibitem[{{Le} {et~al.}(2023){Le}, {Stanier}, {Yin}, {Wetherton}, {Keenan}, \& {Albright}}]{Le23}
{Le}, A., {Stanier}, A., {Yin}, L., {et~al.} 2023, Physics of Plasmas, 30, 063902, \dodoi{10.1063/5.0146529}

\bibitem[{{Le Contel} {et~al.}(2016){Le Contel}, {Retin{\`o}}, {Breuillard}, {Mirioni}, {Robert}, {Chasapis}, {Lavraud}, {Chust}, {Rezeau}, {Wilder}, {Graham}, {Argall}, {Gershman}, {Lindqvist}, {Khotyaintsev}, {Marklund}, {Ergun}, {Goodrich}, {Burch}, {Torbert}, {Needell}, {Chutter}, {Rau}, {Dors}, {Russell}, {Magnes}, {Strangeway}, {Bromund}, {Leinweber}, {Plaschke}, {Fischer}, {Anderson}, {Le}, {Moore}, {Pollock}, {Giles}, {Dorelli}, {Avanov}, \& {Saito}}]{LeContel16}
{Le Contel}, O., {Retin{\`o}}, A., {Breuillard}, H., {et~al.} 2016, \grl, 43, 5943, \dodoi{10.1002/2016GL068968}

\bibitem[{{Lindqvist} {et~al.}(2016){Lindqvist}, {Olsson}, {Torbert}, {King}, {Granoff}, {Rau}, {Needell}, {Turco}, {Dors}, {Beckman}, {Macri}, {Frost}, {Salwen}, {Eriksson}, {{\AA}hl{\'e}n}, {Khotyaintsev}, {Porter}, {Lappalainen}, {Ergun}, {Wermeer}, \& {Tucker}}]{Lindqvist16}
{Lindqvist}, P.-A., {Olsson}, G., {Torbert}, R.~B., {et~al.} 2016, \ssr, 199, 137, \dodoi{10.1007/s11214-014-0116-9}

\bibitem[{Liu {et~al.}(2022)Liu, Vaivads, Khotyaintsev, Fu, Graham, Steinvall, Liu, \& Burch}]{liu2022cross}
Liu, C., Vaivads, A., Khotyaintsev, Y.~V., {et~al.} 2022, The Astrophysical Journal, 926, 198

\bibitem[{{Malaspina} {et~al.}(2024){Malaspina}, {Ergun}, {Cairns}, {Short}, {Verniero}, {Cattell}, \& {Livi}}]{malaspina2024frequency}
{Malaspina}, D.~M., {Ergun}, R.~E., {Cairns}, I.~H., {et~al.} 2024, \apj, 969, 60, \dodoi{10.3847/1538-4357/ad4b12}

\bibitem[{{Malaspina} {et~al.}(2018){Malaspina}, {Ukhorskiy}, {Chu}, \& {Wygant}}]{Malaspina18}
{Malaspina}, D.~M., {Ukhorskiy}, A., {Chu}, X., \& {Wygant}, J. 2018, \jgr, 123, 2566, \dodoi{10.1002/2017JA025005}

\bibitem[{{Masuda} {et~al.}(1994){Masuda}, {Kosugi}, {Hara}, {Tsuneta}, \& {Ogawara}}]{Masuda94}
{Masuda}, S., {Kosugi}, T., {Hara}, H., {Tsuneta}, S., \& {Ogawara}, Y. 1994, \nat, 371, 495, \dodoi{10.1038/371495a0}

\bibitem[{Matteini {et~al.}(2010)Matteini, Landi, Velli, \& Hellinger}]{matteini2010kinetics}
Matteini, L., Landi, S., Velli, M., \& Hellinger, P. 2010, Journal of Geophysical Research: Space Physics, 115

\bibitem[{Medvedev {et~al.}(1998)Medvedev, Diamond, Rosenbluth, \& Shevchenko}]{medvedev1998asymptotic}
Medvedev, M., Diamond, P., Rosenbluth, M., \& Shevchenko, V. 1998, Physical review letters, 81, 5824

\bibitem[{Moore {et~al.}(2016)Moore, Nykyri, \& Dimmock}]{moore2016cross}
Moore, T., Nykyri, K., \& Dimmock, A. 2016, Nature Physics, 12, 1164

\bibitem[{Mozer {et~al.}(2020)Mozer, Bonnell, Bowen, Schumm, \& Vasko}]{mozer2020large}
Mozer, F., Bonnell, J., Bowen, T., Schumm, G., \& Vasko, I. 2020, The Astrophysical Journal, 901, 107

\bibitem[{{Pollock} {et~al.}(2016){Pollock}, {Moore}, {Jacques}, {Burch}, {Gliese}, {Saito}, {Omoto}, {Avanov}, {Barrie}, {Coffey}, {Dorelli}, {Gershman}, {Giles}, {Rosnack}, {Salo}, {Yokota}, {Adrian}, {Aoustin}, {Auletti}, {Aung}, {Bigio}, {Cao}, {Chandler}, {Chornay}, {Christian}, {Clark}, {Collinson}, {Corris}, {De Los Santos}, {Devlin}, {Diaz}, {Dickerson}, {Dickson}, {Diekmann}, {Diggs}, {Duncan}, {Figueroa-Vinas}, {Firman}, {Freeman}, {Galassi}, {Garcia}, {Goodhart}, {Guererro}, {Hageman}, {Hanley}, {Hemminger}, {Holland}, {Hutchins}, {James}, {Jones}, {Kreisler}, {Kujawski}, {Lavu}, {Lobell}, {LeCompte}, {Lukemire}, {MacDonald}, {Mariano}, {Mukai}, {Narayanan}, {Nguyan}, {Onizuka}, {Paterson}, {Persyn}, {Piepgrass}, {Cheney}, {Rager}, {Raghuram}, {Ramil}, {Reichenthal}, {Rodriguez}, {Rouzaud}, {Rucker}, {Saito}, {Samara}, {Sauvaud}, {Schuster}, {Shappirio}, {Shelton}, {Sher}, {Smith}, {Smith}, {Smith}, {Steinfeld}, {Szymkiewicz}, {Tanimoto}, {Taylor}, {Tucker}, {Tull}, {Uhl}, {Vloet}, {Walpole},
  {Weidner}, {White}, {Winkert}, {Yeh}, \& {Zeuch}}]{Pollock16:mms}
{Pollock}, C., {Moore}, T., {Jacques}, A., {et~al.} 2016, \ssr, 199, 331, \dodoi{10.1007/s11214-016-0245-4}

\bibitem[{Raouafi {et~al.}(2016)Raouafi, Patsourakos, Pariat, Young, Sterling, Savcheva, Shimojo, Moreno-Insertis, DeVore, Archontis, {et~al.}}]{raouafi2016solar}
Raouafi, N., Patsourakos, S., Pariat, E., {et~al.} 2016, Space Science Reviews, 201, 1

\bibitem[{Roberts(2006)}]{roberts2006slow}
Roberts, B. 2006, Philosophical Transactions of the Royal Society A: Mathematical, Physical and Engineering Sciences, 364, 447

\bibitem[{{Runov} {et~al.}(2009){Runov}, {Angelopoulos}, {Sitnov}, {Sergeev}, {Bonnell}, {McFadden}, {Larson}, {Glassmeier}, \& {Auster}}]{Runov09grl}
{Runov}, A., {Angelopoulos}, V., {Sitnov}, M.~I., {et~al.} 2009, \grl, 36, L14106, \dodoi{10.1029/2009GL038980}

\bibitem[{{Russell} {et~al.}(2016){Russell}, {Anderson}, {Baumjohann}, {Bromund}, {Dearborn}, {Fischer}, {Le}, {Leinweber}, {Leneman}, {Magnes}, {Means}, {Moldwin}, {Nakamura}, {Pierce}, {Plaschke}, {Rowe}, {Slavin}, {Strangeway}, {Torbert}, {Hagen}, {Jernej}, {Valavanoglou}, \& {Richter}}]{Russell16:mms}
{Russell}, C.~T., {Anderson}, B.~J., {Baumjohann}, W., {et~al.} 2016, \ssr, 199, 189, \dodoi{10.1007/s11214-014-0057-3}

\bibitem[{Sagdeev(1979)}]{sagdeev19791976}
Sagdeev, R.~Z. 1979, Reviews of Modern Physics, 51, 1

\bibitem[{Smith \& Priest(1972)}]{smith1972current}
Smith, D.~F., \& Priest, E. 1972, Astrophysical Journal, vol. 176, p. 487, 176, 487

\bibitem[{{Sonnerup} \& {Cahill}(1968)}]{Sonnerup68}
{Sonnerup}, B.~U.~{\"O}., \& {Cahill}, Jr., L.~J. 1968, \jgr, 73, 1757, \dodoi{10.1029/JA073i005p01757}

\bibitem[{{Sonnerup} \& {Scheible}(2000)}]{bookISSI:Sonnerup}
{Sonnerup}, B.~U.~{\"O}., \& {Scheible}, M. 2000, ESA Special Publication, Vol. 449, ISSI Book on Analysis Methods for Multi-Spacecraft Data, ed. {G.Paschmann and Patrick W. D. }

\bibitem[{Sorriso-Valvo {et~al.}(2019)Sorriso-Valvo, Catapano, Retin{\`o}, Le~Contel, Perrone, Roberts, Coburn, Panebianco, Valentini, Perri, {et~al.}}]{sorriso2019turbulence}
Sorriso-Valvo, L., Catapano, F., Retin{\`o}, A., {et~al.} 2019, Physical Review Letters, 122, 035102

\bibitem[{Stawarz {et~al.}(2016)Stawarz, Eriksson, Wilder, Ergun, Schwartz, Pouquet, Burch, Giles, Khotyaintsev, Contel, {et~al.}}]{stawarz2016observations}
Stawarz, J., Eriksson, S., Wilder, F., {et~al.} 2016, Journal of Geophysical Research: Space Physics, 121, 11

\bibitem[{{Ukhorskiy} {et~al.}(2022){Ukhorskiy}, {Sorathia}, {Merkin}, {Crabtree}, {Fletcher}, {Malaspina}, \& {Schwartz}}]{Ukhorskiy22:NatSR}
{Ukhorskiy}, A.~Y., {Sorathia}, K.~A., {Merkin}, V.~G., {et~al.} 2022, Scientific Reports, 12, 4446, \dodoi{10.1038/s41598-022-08038-x}

\bibitem[{{Valentini} {et~al.}(2011){Valentini}, {Perrone}, \& {Veltri}}]{valentini2011short}
{Valentini}, F., {Perrone}, D., \& {Veltri}, P. 2011, \apj, 739, 54, \dodoi{10.1088/0004-637X/739/1/54}

\bibitem[{Valentini {et~al.}(2014)Valentini, Vecchio, Donato, Carbone, Briand, Bougeret, \& Veltri}]{valentini2014nonlinear}
Valentini, F., Vecchio, A., Donato, S., {et~al.} 2014, The Astrophysical Journal Letters, 788, L16

\bibitem[{Valentini {et~al.}(2008)Valentini, Veltri, Califano, \& Mangeney}]{valentini2008cross}
Valentini, F., Veltri, P., Califano, F., \& Mangeney, A. 2008, Physical review letters, 101, 025006

\bibitem[{{Vasko} {et~al.}(2022){Vasko}, {Mozer}, {Bale}, \& {Artemyev}}]{Vasko22:grl}
{Vasko}, I.~Y., {Mozer}, F.~S., {Bale}, S.~D., \& {Artemyev}, A.~V. 2022, \grl, 49, e98640, \dodoi{10.1029/2022GL098640}

\bibitem[{{Vidal-Luengo} {et~al.}(2025){Vidal-Luengo}, {Malaspina}, \& {Eriksson}}]{vidal2025search}
{Vidal-Luengo}, S.~E., {Malaspina}, D.~M., \& {Eriksson}, S. 2025, \apjl, 991, L14, \dodoi{10.3847/2041-8213/adff88}

\bibitem[{{Wang} {et~al.}(2020){Wang}, {Vasko}, {Mozer}, {Bale}, {Artemyev}, {Bonnell}, {Ergun}, {Giles}, {Lindqvist}, {Russell}, \& {Strangeway}}]{Wang20:apjl}
{Wang}, R., {Vasko}, I.~Y., {Mozer}, F.~S., {et~al.} 2020, \apjl, 889, L9, \dodoi{10.3847/2041-8213/ab6582}

\bibitem[{Wilder {et~al.}(2016)Wilder, Ergun, Schwartz, Newman, Eriksson, Stawarz, Goldman, Goodrich, Gershman, Malaspina, {et~al.}}]{wilder2016observations}
Wilder, V., Ergun, R., Schwartz, S., {et~al.} 2016, Geophysical Research Letters, 43, 8859

\bibitem[{{Woodham} {et~al.}(2021){Woodham}, {Horbury}, {Matteini}, {Woolley}, {Laker}, {Bale}, {Nicolaou}, {Stawarz}, {Stansby}, {Hietala}, {Larson}, {Livi}, {Verniero}, {McManus}, {Kasper}, {Korreck}, {Raouafi}, {Moncuquet}, \& {Pulupa}}]{woodham2021enhanced}
{Woodham}, L.~D., {Horbury}, T.~S., {Matteini}, L., {et~al.} 2021, \aap, 650, L1, \dodoi{10.1051/0004-6361/202039415}

\bibitem[{Zanelli {et~al.}(2025)Zanelli, Perri, Condoluci, Veltri, Pegoraro, Pezzi, Perrone, Trotta, \& Valentini}]{zanelli2025flat}
Zanelli, S., Perri, S., Condoluci, M., {et~al.} 2025, Physics of Plasmas, 32

\end{thebibliography}

\bibliographystyle{aasjournal}



\end{document}